# High-speed readout for direct light orbital angular momentum photodetector via photoelastic modulation

Dehong Yang[1], Chang Xu[1], Jiawei Lai[2], Zipu Fan[1], Delang Liang[1,3], Shiyu Wang[1], Jinluo Cheng[4], Dong Sun[1,5,6,*]

[1]International Center for Quantum Materials, School of Physics, Peking University, Beijing, China
[2]Ministry of Education Key Laboratory for Nonequilibrium Synthesis and Modulation of Condensed Matter, Shaanxi Province Key Laboratory of Quantum Information and Quantum Optoelectronic Devices, School of Physics, Xi'an Jiaotong University, Xi'an, China
[3]Key Laboratory for Micro-Nano Physics and Technology of Hunan Province, Hunan Institute of Optoelectronic Integration, College of Materials Science and Engineering, Hunan University, Changsha, China
[4]GPL Photonics Laboratory, State Key Laboratory of Luminescence Science and Technology, Changchun Institute of Optics, Fine Mechanics and Physics, Chinese Academy of Sciences, Changchun, China
[5]Collaborative Innovation Center of Quantum Matter, Beijing, China
[6]Frontiers Science Center for Nano-optoelectronics, School of Physics, Beijing, China
*Email: sundong@pku.edu.cn;

**Abstract.** Recent progress in direct photodetection of light orbital angular momentum (OAM) based on the orbital photogalvanic effect (OPGE) provides an effective way for on-chip direct electric readout of orbital angular momentum, as well as large-scale integration focal-plane array devices. However, the recognition of OAM order from photocurrent response requires the extraction of circular polarization-dependent response. To date, the operation speed of such detector is currently at the minute level and is limited by slow mechanical polarization modulation and low OAM recognition capability. In this work, we demonstrate that the operation speed can be greatly improved via electrical polarization modulation strategy with photoelasitc modulator accompanied by phase-locked readout approach with lock-in amplifier. We demonstrate an operation speed of up to 1 kHz with this new technology in the mid-infrared region (4 μm) on an OAM detector using multilayer graphene (MLG) as photosensitive material. In principle, with new modulation and readout scheme, we can potentially increase the operation speed to 50.14 kHz with a PEM that operates at a state-of-the-art speed. Our work paves the way toward high-speed operation of direct OAM detection devices based on OPGE effect and pushes such technology to a more practical stage for focal plane array applications.

*Dong Sun, Email: sundong@pku.edu.cn;

## 1 Introduction

Recent developments in light orbital angular momentum (OAM) detectors based on the orbital photogalvanic effect (OPGE) of semimetallic materials provide a promising route toward on-chip direct electric readout of the OAM of light, as well as large-scale integration of focal-plane array OAM detection devices[1-3]. Compared with the parallel technology route of on-chip OAM detection, which is based on surface plasmon polaritons (SPPs), OPGE-based OAM detection has a broader operation wavelength range, higher responsivity and simpler device structure[4-8]. This OAM detection scheme is based on the orbital photogalvanic effect (OPGE) driven by the helical phase gradient of light through the electric quadrupole and magnetic dipole response of materials, which is related to the fourth-rank nonlinear tensor of the detection material. The crystal symmetry of the detection material needs to fulfill the symmetry requirement to have a nonvanishing OPGE response. The derivation of the expression for the OPGE response have been described in detail in Refs. [1-3]. In general, under the excitation of a Laguerre Gaussian (LG) beam carrying OAM, the photocurrent response arises from the electric quadrupole and magnetic dipole effects corresponding to the response term $J_{qp}$, which can be divided into four terms according to its dependence on the SAM ($\sigma_i$) and OAM ($m$) as follows:

$$J_{qp}(\rho,\theta,z) = m \cdot \sigma_i J_{(1)}(\rho,\theta,z) + m J_{(2)}(\rho,\theta,z) + \sigma_i J_{(3)}(\rho,\theta,z) + J_{(4)}(\rho,\theta,z) \quad (1)$$

The first term $m \cdot \sigma_i J_{(1)}(\rho,\theta,z)$ is proportional to the product of SAM ($\sigma_i$) and OAM (m) and can be used for OAM detection. Experimentally, the OAM-dependent photocurrent component must be extracted experimentally from a circular polarization-dependent measurement, commonly denoted as the circular photogalvanic effect (CPGE), and the extracted component ($m \cdot J_{(1)}$) *is proportional to the OAM order m if the power and ring radius of LG beams remain the same for different OAM orders*[1,2]. Then, the topological charge of light OAM can be directly distinguished by the quantized plateau of the CPGE component.

However, to extract the CPGE response, the incoming light must be modulated between the left and right circular polarizations, which was realized previously through continuously rotating a quarter waveplate (QWP), as illustrated in Fig. 1a[1-3]. The CPGE component is then extracted via Fourier transform of the QWP angle-dependent photocurrent response. Limited by the speed of mechanical polarization modulation, the operation speed of all previous works is at the minute level, which cannot fulfill the speed requirements of most applications[9-14]. Currently, the slow operation speed is the major drawback of OPGE-based OAM detectors compared with parallel OAM detection technology based on SPPs, which can reach operation speeds on the order of tens of microseconds[4].

To achieve high-speed direct detection of OAM, the key is to renovate the polarization modulation technology to increase the circular polarization modulation speed and supplement it with a fast readout technique to extract the CPGE component from the modulated response simultaneously. For fast circular polarization modulation, the use of high-speed electrical polarization modulation techniques, such as photoelastic modulators (PEMs) and electro-optic modulators (EOMs), to replace traditional mechanical modulation could promote the polarization modulation speed up to the MHz and GHz levels for PEMs and EOMs, respectively[15-20]. The generated polarization-modulated photocurrent response can be extracted via a phase-sensitive detection technique with a lock-in amplifier locked to the polarization modulation frequency, as illustrated in Fig. 1b. In this work, we demonstrate this electric modulation scheme experimentally via the PEM-based polarization modulation method to replace traditional mechanical modulation to achieve a 50-kHz modulation speed, and the fast-modulated CPGE response can be directly extracted from a lock-in amplifier that is locked to the modulation signal of the PEM. The fast electrical modulation scheme is demonstrated on an OPGE-based OAM detector made from multilayer graphene (MLG). The topological charge of light OAM can be clearly distinguished by the quantized plateau of the CPGE response, which is directly extracted by a lock-in amplifier. Moreover, we compare the OPGE responsivity and OAM resolution capability under mechanical and PEM modulations. We find that the difference in the OPGE responsivity arises from the different polarization modulation and signal readout techniques. Considering that the polarization modulation frequency of the PEM is 50.14 kHz, the mini-second operation speed of the detector is jointly limited by the requirement of the integration time of multiple modulation periods in the readout process using a lock-in amplifier and the microsecond-level response time of the device. Our work solves the major obstacle of slow operation speed caused by the traditional mechanical polarization modulation technique in OPGE detectors, and the new modulation and readout scheme is directly applicable for large-scale integration of a focal plane array device[21-25].

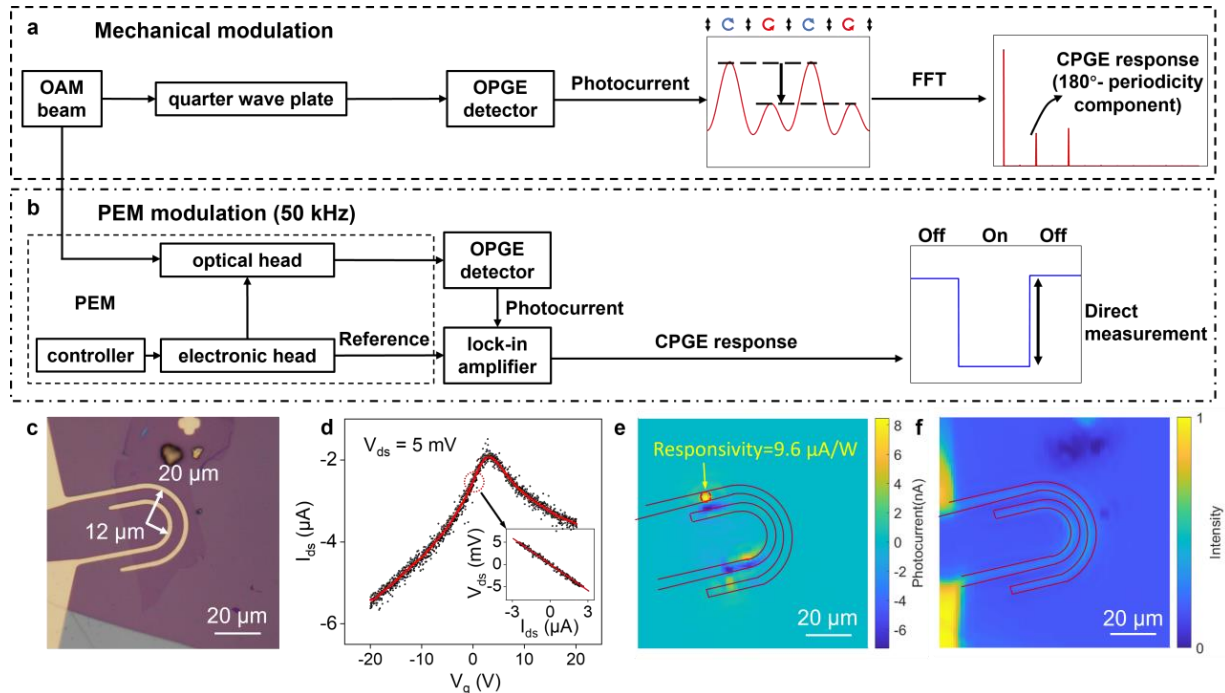

**Fig. 1** Diagrams of different polarization modulation and readout schemes for the OPGE detector and basic characterization of the MLG photodetector: (a-b) diagrams of OAM photodetection for (a) mechanical modulation and (b) PEM modulation. (c) Optical image of the MLG photodetector. (d) Drain-source current $I_{ds}$ as a function of the back gate voltage $V_g$ with a 5-mV drain-source voltage $V_{ds}$, together with $I_{ds}$-$V_{ds}$ measurements with zero back gate voltage in the inset. (e-f) Scanning photocurrent mapping (e) together with in situ scanning reflection mapping (f) under the excitation of a basic mode Gaussian beam with a power of 0.9 mW at 4 μm. The spatial resolution is approximately 10 μm.

## 2 MLG OAM Detector and OPGE Response

### 2.1 OAM detector device and basic characterization

The direct OAM detector used in this work is made with few-layer exfoliated graphene, as shown in Fig. 1c. The graphene flakes are exfoliated from the graphite and transferred onto a 300 nm/500 μm SiO$_2$/Si substrate. A standard electron-beam lithography technique is used to pattern the U-shaped electrodes to collect the radial photocurrent response of the MLG, as the OPGE response survives along the radial direction according to the symmetry of the MLG[1]. Then, the electrodes were deposited by an electron-beam evaporator with 10 nm Ti and 80 nm Au. The deviation of the OPGE response of the MLG device is fully described in Supplementary Section 1. Compared with TaIrTe$_4$, which has also been demonstrated to realize OAM detection in the mid-infrared region[2], MLG has one order of magnitude greater OPGE responsivity and recognition capability[1]. Moreover, MLGs are already epitaxially growable at the wafer scale through either chemical vapor deposition or epitaxial growth on SiC, and the fabrication process is completely CMOS compatible, which promises large-scale integration of ambient stable, mid-infrared direct OAM photodetection devices and OAM-sensitive focal plane array devices.

For the specific device used in this work, the radii of the inner and outer electrodes are 12 μm and 20 μm, respectively. Fig. 1d shows the source–drain current $I_{ds}$ as a function of the back gate voltage $V_g$ under a source–drain voltage $V_{ds}$=5 mV, together with the $I_{ds}$-$V_{ds}$ measurement at $V_g$=0. In the measurement, the back gate voltage $V_g$ is applied by a low-resistivity silicon substrate. The peak in the $I_{ds}$-$V_g$ plot signifies the Fermi level at the Dirac point, whereas the linear result from the $I_{ds}$-$V_{ds}$ measurement (inset of Fig. 1d) confirms good ohmic contact. For scanning photocurrent measurement, the output from a 4-μm CW quantum cascade laser source is focused into a spot with a radius of 10 μm by a 40X reflection objective, and scanning microscopy is obtained by controlling the 2D (x,y) movement of the device placed on a motorized stage. Fig. 1e and Fig. 1f show the scanning photocurrent microscopy image of the U-shaped MLG detector under the excitation of a basic mode Gaussian beam at 4 μm with a power of 0.7 mW, together with the in situ scanning reflection image. The photocurrent response mainly originates from the region near the electrodes, with a responsivity lower than that of the previously reported device because of the reduced thickness[1]. Typically, OAM beams are generated by passing the laser beam through a spiral phase plate specially designed for different OAM orders at 4 μm. Then, the OAM beams are focused to the same ring radius and embedded between the inner and outer electrodes. The radial photocurrent can be effectively collected by the U-shaped electrode, and the topological charge of light OAM can be clearly distinguished by the quantized plateau of the CPGE component of the radial OPGE response, which enables the detection of light OAM.

## 2.2 OPGE response measured with mechanical polarization modulation

First, we present the measurement results of the CPGE response via conventional mechanical modulation. For such measurements, a quarter wave plate (QWP) is placed after a polarizer and rotated to modulate the polarization of the OAM beams following the scheme shown in Fig. 2a When the QWP angle ($\theta = 2\pi f t$) is rotated with rotation frequency f, the polarization state of the OAM beams undergoes a 180° periodic change from linear ($\theta = 0°$)-left circular ($\theta = 45°$)-linear ($\theta = 90°$)-right circular ($\theta = 135°$)-linear ($\theta = 180°$), as shown in Fig. 2b. The dependence of the first term of the radial OPGE response $m \cdot \sigma_i J^\rho_{(1)}$ on the QWP angle $\theta$ (or $2\pi f t$) can be written as:

$$m \cdot \sigma_i J^\rho_{(1)} = -m \cdot C(\vec{r}) \sin 4\pi f t \tag{2}$$

where $C(\vec{r})$ is a coefficient related to the light field distribution and the rank-4 conductivity tensors of the detection material. The spatial integration of $C(\vec{r})$, which corresponds to the collected radial OPGE response, is a function of the total power $P$, the ring radius of the focused OAM beam and the rank-4 conductivity tensors of the detection material[1]. The deviation of Equation (2) is fully described in Supplementary Section 2. According to Equation (2), by measuring the photocurrent response at different QWP angles ($\theta$), the CPGE component can be obtained by extracting the 180°-periodic component of the photocurrent response through a Fourier transform, and the extracted CPGE response is written as $m \cdot C(\vec{r})$, which is proportional to the OAM order m. Previous work on OAM photodetectors based on WTe$_2$, TaIrTe$_4$ and MLG employed these mechanical polarization modulation and CPGE extraction approaches[1-3]. The topological charge of light OAM can be clearly distinguished by the quantized plateau of the CPGE response, but the operation speed is limited to the order of minutes[1-3].

The CPGE measurement results with mechanical modulation are presented in Fig. 2c-e, which are qualitatively the same as the results presented in Ref. [1]; however, the measurements in this work are performed on a thinner MLG device for fair comparison between the two modulation schemes. In the measurement, the OAM beams are focused by the 40X reflection objective to the ring with a radius of 16 μm and embedded between the inner and outer U-shaped electrodes. Fig. 2c shows the photocurrent as a function of the angle of the QWP for OAM orders of ±4, ±2 and ±1 with an excitation power of 1.5 mW. The difference in the photocurrent response for excitations with left and right circularly polarized beams is marked in the figure, which shows similar magnitudes but opposite signs for the OAM order ±m, and the magnitude increases as |m| increases. Fig. 2d shows the extracted CPGE component J$_C$ as a function of OAM order m. The J$_C$ shows step-like changes with OAM order m and is proportional to m without a background signal. These results are consistent with a previous report on a similar MLG device[1]. From the measurement results, we can obtain the major merits of device performance. If we define the OPGE responsivity $K$ as the ratio of the CPGE component $J_C$ to the OAM order $m$ under unit OAM light excitation power ($P$): $K = J_C/m \cdot P$. Linear fitting with respect to m in Fig. 2e can be used to obtain the responsivity ($K$) of the OPGE and the uncertainty of the responsivity $K$ ($\sigma_K$) to be K=74.9 nA/W and $\sigma_K$ =4 nA/W. Subsequently, we define the resolution capability $R$ of OAM as $R = K/\sigma_K$ to account for the signal-to-noise ratio of the OPGE measurement, and we can obtain R=18.6. Although the OPGE responsivity of this device is half of that previously reported for a thicker MLG device, these two devices have comparable OAM resolution capabilities because of the reduced noise level of the thinner device used in this work[1]. Nevertheless, the performance of the device used in this work is still five times better than that demonstrated for a TaIrTe$_4$ device[2].

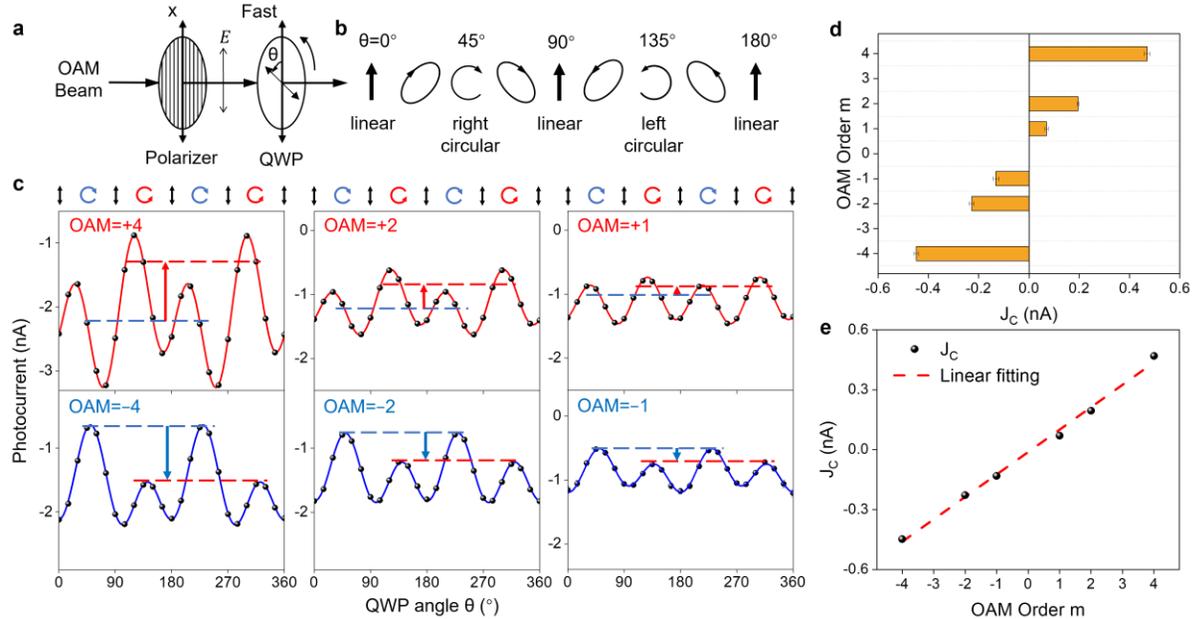

**Fig. 2** OPGE response of the MLG device based on mechanical modulation: (a) Polarization modulation scheme based on mechanically rotating a quarter wave plate; (b) Polarization modulation as a function of the QWP angle $\theta$ in a period of 180°; (c) Photocurrent response of the device as a function of $\theta$; the PC responses for the left and right circular polarizations are marked by blue and red dashed lines, respectively. The CPGE component is marked by arrows, with red and blue representing positive and negative CPGE responses, respectively. (d) CPGE component $J_C$ as a function of the OAM order $m$. The error bars represent the standard deviation of the fit. (e) Linear fitting of $J_C$ as a function of the OAM order $m$.

## 2.3 OPGE response measured with photoelastic modulation

Next, we present the measurement results when a photoelastic modulator (PEM) is used to replace the QWP to achieve high-speed electric polarization modulation at a frequency of 50.14 kHz and directly extract the CPGE component of the photocurrent response via the lock-in amplifier. A detailed schematic diagram of the measurement setup is shown in Fig. 3a-b. The PEM consists of a piezoelectric transducer and a half wave resonant bar[26,27]. The transducer changes the birefringence properties of the resonant bar by electrically stretching and compressing the vibration optical element, thereby imparting a periodic phase difference to the polarization components along the two optical axes. In this configuration, the optical axes of the PEM are aligned along the xy direction, the polarizer's polarization direction is set at a 45-degree angle to the x-axis, and the peak phase retardation ($\delta_0$) of the PEM is set to $\pi/2$ (Fig. 3d). To drive the PEM, a sinusoidal modulating voltage (as shown in Fig. 3b) driven by an electronic driver circuit is applied to the quartz piezoelectric transducer to drive the rectangular $ZnSe_2$ optical element and create polarization modulation. The phase difference induced by the PEM for two perpendicular polarizations at different times (t) can be written as $\delta = \delta_0 \sin 2\pi f t$, where $f$ is the modulation frequency. The PEM is a resonant device where the transducer is tuned to the resonance frequency of the optical element, which is determined by its bar length and the speed of sound in the material[26]. Here, the precise oscillation frequency is fixed at 50.14 kHz, which is determined by the photoelastic properties of the $ZnSe_2$ element and transducer assembly in the PEM. When the OAM beam passes through the PEM, in a single operational cycle of the PEM, the polarization of the

OAM beam undergoes a sequence of transitions—linear, left-circular, linear, right-circular, and linear—exhibiting a variation pattern of polarization modulation similar to that realized by rotating the QWP in a period of 180°, as shown in Fig. 3f.

However, the difference in the specific variation in the polarization sequence leads to different variations in the OPGE signal in one operation cycle. For PEM modulation, after the first-order expansion of $\sin\delta$ through the integer-order Bessel function and keeping the leading first-order term, the dependence of the OPGE response term on the QWP angle $\theta = 2\pi ft$ can be written as:

$$m \cdot \sigma_i J^\rho_{(1)} = m \cdot C(\vec{r}) 2 J_1(\delta_0) \sin 2\pi ft \tag{3}$$

where $J_1(\delta_0)$ is a first-order Bessel function. Compared with the mechanical modulation, the CPGE response extracted via the PEM has an additional $2J_1(\delta_0)$ coefficient, and for peak phase retardation $\delta_0 = \pi/2$ used in this work, $2J_1(\pi/2) = 1.13365$. The deviation of Equation (3) is fully described in Supplementary Section 2. Another difference lies in the readout method of the CPGE response. When the PEM is used, the CPGE response can be directly extracted by the lock-in amplifier locked to the 50.14 kHz periodic modulation of the left and right circular polarizations, and the lock-in amplifier outputs the root mean square of the CPGE response. For mechanical modulation, the phase is continuously changed by rotating a quarter waveplate, and the CPGE response is extracted via the Fourier transform of the photocurrent response, which is dependent on the QWP angle. To exclude background signals and suppress 1/f noise, the OAM beams are also modulated by a mechanical chopper, and the photocurrent response is also extracted by a lock-in amplifier locked to the driving frequency of the chopper. The different CPGE read-out approaches introduce additional coefficient differences to the experimentally measured CPGE responses, which are fully described in Supplementary Section 3. Overall, for mechanical and PEM modulations, the extracted CPGE response differs only in terms of the constant coefficient because of different polarization modulations and different CPGE readout methods. Theoretically, the experimentally extracted CPGE response with PEM modulation is $\pi J_1(\pi/2) \approx 1.78$ times greater than that with mechanical modulation. More details about the derivation of the extracted CPGE response are presented in Supplementary Sections S2--3.

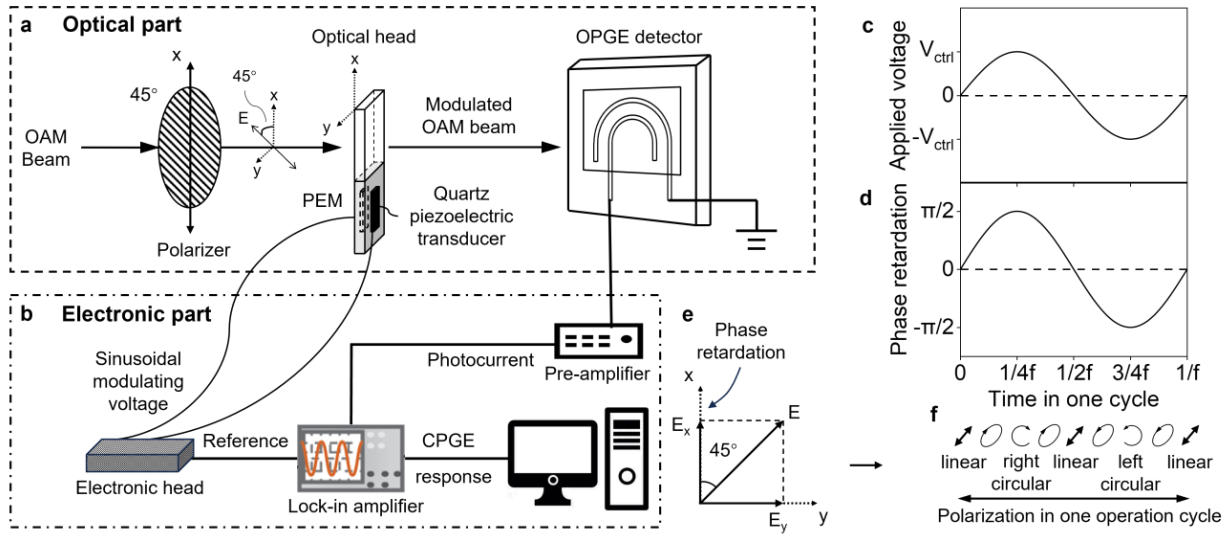

**Fig. 3** Polarization modulation scheme and experimental setup based on the PEM: (a) Schematic diagram of the optical parts (polarization modulation with the polarizer and optical head of the PEM) and (b) the electronic part (modulating voltage, photocurrent collection and CPGE extraction); (c) waveform of the sinusoidal driving voltage applied on the quartz piezoelectric transducer of the PEM; (d) phase retardation introduced by the optical head of the PEM; (e) schematic plot of the polarization direction of incident light and the two principal axis directions of the PEM; (f) polarization modulation in one operation cycle of the PEM.

The measurement results with PEM modulation are presented in Fig. 4. The excitation conditions of the OAM beam are the same as those for the mechanical modulation measurement. Fig. 4a shows the extracted CPGE component $J_C$ at OAM orders of ±4, ±2 and ±1 with a constant excitation power of 1.5 mW. The CPGE signal is measured with a lock-in amplifier that is phase locked to the 50.14-kHz modulation signal of the PEM, and the integration time constant is set to 300 ms, which corresponds to 15000 modulation cycles for the measurement results shown in Fig.

4a. The measurement is performed by comparing the lock-in signal by blocking or unblocking the excitation light. Here, we note that there are clear background signals when the light is turned off. The constant background signal is dominated by the pickup of electromagnetic waves leaked from the electric driving head of the PEM. Since the picked signal has the same frequency as the PEM's operating frequency, it dominates the background that is picked by the lock-in amplifier at the driving frequency. Taking the difference in the on-off signal provides the absolute CPGE response. The results show similar magnitudes but opposite signs for opposite OAM orders ±m, and the magnitude increases as |m| increases. Fig. 4b shows the extracted CPGE component $J_C$ as a function of the OAM order m. The $J_C$ shows step-like changes with the OAM order m and is proportional to m. These results are all consistent with those measured with the mechanical modulation shown in Fig. 2c-e. Here, we note that the magnitude of the extracted CPGE response from PEM modulation is approximately 1.1 times greater than that extracted from mechanical modulation, which is different from the theoretical ratio of 1.78; such deviation is mainly due to the limited response time of the device. The response time for a typical MLG device is measured to be 3.42 μs, as presented in Fig. S1 of the supplementary information. For 50.14 kHz polarization modulation, each modulation cycle is only 20 μs, which is close to the response time of the device, so the response magnitude of the device is reduced because of the limited photocurrent response speed of the devices, which leads to a lower CPGE response magnitude ratio of the two different modulation schemes.

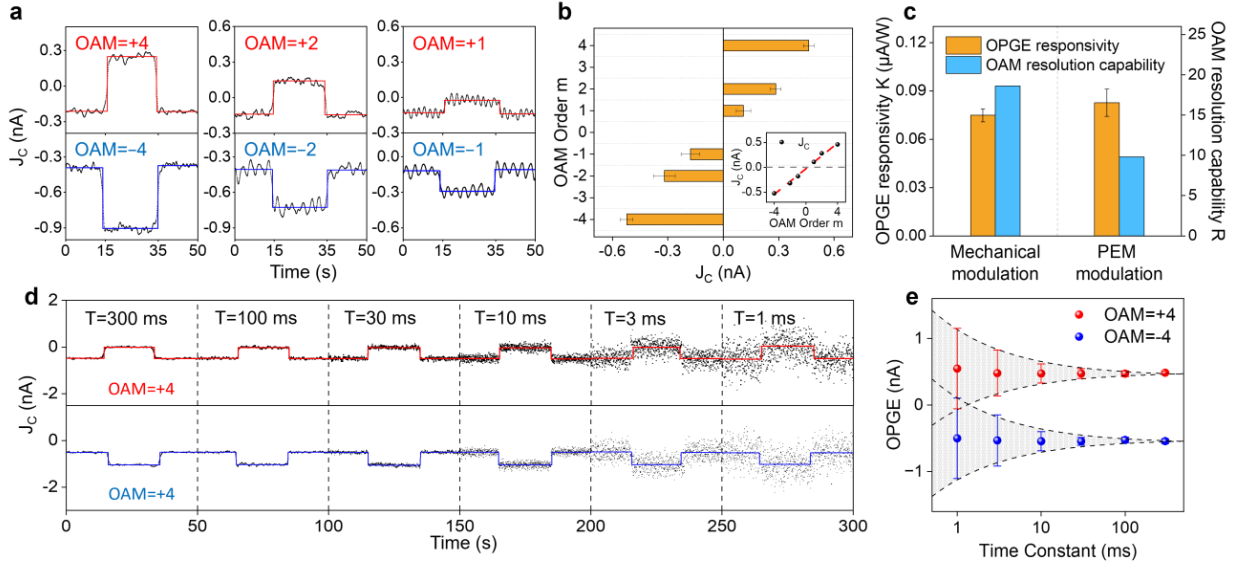

**Fig. 4** OPGE response of the MLG device based on PEM modulation: (a) On-off measurements of the CPGE response under the excitation of OAM beams with OAM orders ±4, ±2, and ±1; (b) CPGE component $J_C$ together with its linear fit as a function of the OAM order $m$ (inset). The error bar is the standard deviation of the fit; (c) Comparison of the OPGE responsivity and OAM resolution capability under mechanical and PEM modulation; (d) On-off measurements of the CPGE response with different time constants of the lock-in amplifier under the excitation of OAM beams with OAM order ±4; (e) Measured CPGE response together with its uncertainty as a function of the time constant of the lock-in amplifier under the excitation of OAM beams with OAM order ±4.

From the measurement results, we can obtain the major merits of the figures of device performance: the OPGE responsivity $K = 82.6$ nA/W, the uncertainty of the responsivity $\sigma_K = 8$ nA/W and the resolution capability $R = K/\sigma_K = 9.8$. Fig. 4c presents a comparison of the OPGE responsivities and resolution capabilities of the two modulation approaches. The OPGE responsivity under PEM modulation is approximately 1.1 times greater than that obtained under mechanical modulation. However, owing to the lower signal-to-noise ratio of the CPGE response, the OAM resolution capability for PEM modulation is approximately one third that for mechanical modulation. The operation speed of PEM modulation is limited mainly by the 300 ms measurement time constant of the lock-in amplifier. The measurement time constant of the lock-in amplifier can be reduced to compensate for the elevated noise level. Fig. 4d shows the measurement results with different measurement time constants (from 1 ms to 300 ms) for m=±4. As the time constant decreases, the photocurrent response remains constant (~0.5 nA), but the noise increases dramatically. When the time constant decreases to 1 ms, the signal can barely be resolved, which limits the final operation speed of the device to 1 kHz. Here, we note the trade-off relationship between the operation speed and the signal–noise ratio plotted in Fig. 4e. However, providing enough OPGE response of the device, the measurement time of the lock-in can be further decreased, and the operation speed can be improved further. The time constant is ultimately limited by the 50.14-kHz polarization modulation frequency of the PEM and the 3.54-μs photocurrent response time of the device. In principle, the minimum lock-in time constant must be over several polarization modulation cycles, and the modulation period must be twice the response time of the device through the Nyquist–Shannon sampling theorem.

## 3 Discussion and perspectives

For the PEM modulation scheme demonstrated in this work, the polarization modulation speed is mainly limited by the resonance frequency of the PEM crystal and the photocurrent response speed of the device. A ZnSe crystal is used in this work to achieve polarization modulation at mid-infrared wavelengths. The resonance frequency of the PEM is determined by the photoelastic properties of the ZnSe element and transducer assembly and varies with the material species and size, which are constrained by specific factors, such as the size of the focal plane array and operation wavelength. In general, the modulation frequency is limited to several MHz with a PEM modulation scheme[19,20]. Alternatively, the operation can be tuned via an electro-optical modulator, which can reach a 100 GHz polarization modulation speed[16,18]. However, compared with PEMs, electro-optical modulators have the drawbacks of a lower numerical aperture, higher driving voltage, and smaller operation wavelength range due to fewer choices of optical materials[27]. The reading speed of the OAM detector, however, is further limited by the phase-sensitive reading part. The noise level and requirement of the signal-to-noise ratio of the device constrain the maximum bandwidth of the low-pass filter used in the lock-in amplifier. This imposes a limit on the minimum time constant of the lock-in amplifier, which restricts the operation speed of the device to multiple polarization modulation periods. On the other hand, the operation speed can be further limited by the photoresponse time of the detector. Once the modulation frequency approaches or exceeds the photocurrent response bandwidth, the response of the device is retarded to the modulated signal, which decreases the overall magnitude of the photocurrent response.

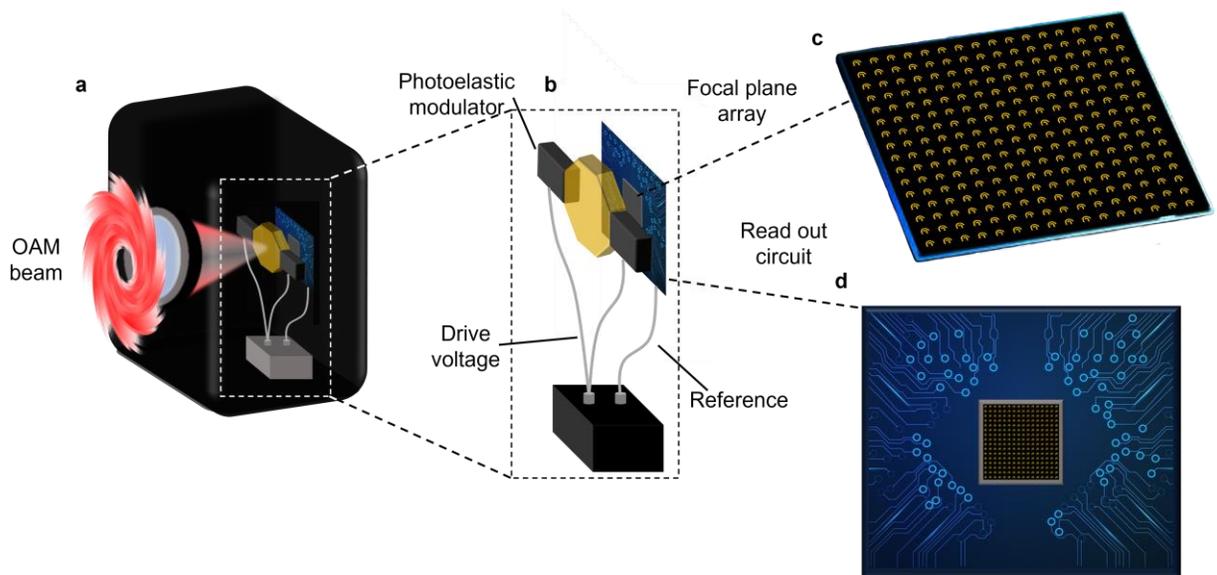

**Fig. 5** Schematic of the application of the PEM modulation scheme for light OAM photodetector focal-plane-array devices based on graphene. (a) The overall device structure of the focal-plane-array device. (b) Schematic of the OAM detection chip and PEM modulation module driven by the power module. (c) Schematic of the focal-plane array based on MLG photodetectors. (d) Schematic of the read-out circuit.

Furthermore, the PEM modulation scheme demonstrated in this work is directly applicable to focal plane array devices, as illustrated in the schematic diagram proposed in Fig. 5. The OAM detector based on MLG can be scaled to focal plane arrays with large areas epitaxially or CVD-grown multilayer graphene[28-30]. The photoelastic crystal with the transducer assembly can be placed before the OAM detector arrays, and the phase-sensitive reading circuit can be integrated with the photocurrent readout circuit and locked to the electric driving source to extract the CPGE component of each detector cell. According to the extracted CPGE component of each sensing cell, the OAM order can be extracted directly to form the desired OAM image from the focal plane arrays. For on-chip integration, miniaturized electrically driven polarization modulation materials are required for OAM detectors, recent progress in the use of atomically thin topological semimetals for tunable phase retardation may satisfy such demand[31], and the driving voltage can be much lower for atomically thin materials and thus facilitate the on-chip integration of all necessary components. The 3-4 orders improvement in the operation speed demonstrated in this work sweeps away the major technical bottleneck of OPGE-based OAM photodetectors, which extends the application scope of such OAM detectors into a new era.

*Disclosures*

The authors declare no competing interests.

*Code, Data, and Materials Availability*

All data supporting the findings of this study are available within the article and its Supplementary Information and via the figshare repository [link.]. Further datasets are available from the corresponding author upon reasonable request.

*Acknowledgments*

This project was supported by the National Natural Science Foundation of China (Grant Nos. 62250065), the National Key Research and Development Program of China (Grant Nos.

2021YFA1400100 and 2020YFA0308800), the authors also would like to thank the support from the National Natural Science Foundation of China (Grant No: 12404389, 12034001, 62325401, 12034003 and 62227822), the Natural Science Basic Research Program of Shaanxi (Program No. 2024JC-YBQN-0063), and the Open Fund of the State Key Laboratory of Infrared Physics (Grant No. SITP-NLIST-ZD-2023-02).

**Caption List**

**Fig. 1** Diagrams of OAM photodetection and basic characterization of the MLG photodetector: (a-b) diagrams of OAM photodetection for (a) mechanical modulation and (b) PEM modulation. (c) Optical image of the MLG photodetector. (d) Drain-source current $I_{ds}$ as a function of the back gate voltage $V_g$ with a 5 mV drain-source voltage $V_{ds}$, together with $I_{ds}$-$V_{ds}$ measurements with zero back gate voltage. (e-f) Scanning photocurrent mapping (e) together with in situ scanning reflection mapping (f) under the excitation of a basic mode Gaussian beam with a power of 0.9 mW at 4 μm. The spatial resolution is approximately 10 μm.

**Fig. 2** OPGE response of the MLG device based on mechanical modulation: (a) Schematic of polarization modulation based on a polarizer and a quarter wave plate. (b) Polarization modulation as a function of the QWP angle over a period of 180°. (c) Photocurrent response of the device as a function of the quarter-wave plate angle $\theta$. The PC responses for the left and right circular polarizations are marked by blue and red dashed lines, respectively, and the CPGE component is marked by arrows, with red and blue representing positive and negative CPGE responses, respectively. (d) CPGE component $J_C$ as a function of the OAM order $m$. The error bars represent the standard deviation of the fit. (e) Linear fitting of $J_C$ as a function of the OAM order $m$.

**Fig. 3** Schematic of OAM photodetection based on PEM modulation: (a) Optical part (polarization modulation with a polarizer and an optical head of the PEM). (b) Electronic part (modulating

voltage, photocurrent collection and CPGE extraction). (c) Sinusoidal modulating voltage applied on the quartz piezoelectric transducer. (d) Phase retardation introduced by the optical head of the PEM. (e) Schematic for polarization modulation in the two principal axis directions of the PEM. (f) Polarization modulation in one operation cycle of the PEM.

**Fig. 4** OPGE response of the MLG device based on PEM modulation: (a) On-off measurements of the CPGE response under the excitation of OAM beams with OAM orders $\pm 4$, $\pm 2$, and $\pm 1$. (b) CPGE component $J_C$ together with its linear fit as a function of the OAM order $m$. The error bars represent the standard deviation of the fit. (c) Comparison of the OPGE responsivity and OAM resolution capability under mechanical and PEM modulations. (d) On-off measurements of the CPGE response with different time constants of the lock-in amplifier under the excitation of OAM beams with an OAM order of $\pm 4$. (e) Measured CPGE response together with its uncertainty as a function of the time constant of the lock-in amplifier under the excitation of OAM beams with OAM order $\pm 4$.

**Fig. 5** Schematic of the PEM modulation scheme for a light OAM photodetector focal-plane-array device based on graphene. (a) The overall device structure of the focal-plane-array device. (b) Schematic of the OAM detection chip and PEM modulation module driven by the power module. (c) Schematic of the focal-plane array based on MLG photodetectors. (d) Schematic of the read-out circuit.

# Supplementary information
## for
## High-speed readout for direct light orbital angular momentum photodetector via photoelastic modulation

Dehong Yang[1], Chang Xu[1], Jiawei Lai[2], Zipu Fan[1], Delang Liang[1,3], Shiyu Wang[1], Jinluo Cheng[4†], Dong Sun[1,5,6†]

[1]International Center for Quantum Materials, School of Physics, Peking University, Beijing, China

[2]Ministry of Education Key Laboratory for Nonequilibrium Synthesis and Modulation of Condensed Matter, Shaanxi Province Key Laboratory of Quantum Information and Quantum Optoelectronic Devices, School of Physics, Xi'an Jiaotong University, Xi'an, China

[3]Key Laboratory for Micro-Nano Physics and Technology of Hunan Province, Hunan Institute of Optoelectronic Integration, College of Materials Science and Engineering, Hunan University, Changsha, China

[4]GPL Photonics Laboratory, State Key Laboratory of Luminescence Science and Technology, Changchun Institute of Optics, Fine Mechanics and Physics, Chinese Academy of Sciences, Changchun, China

[5]Collaborative Innovation Center of Quantum Matter, Beijing, China

[6]Frontiers Science Center for Nano-optoelectronics, School of Physics, Beijing, China

†Email: sundong@pku.edu.cn;

**Table of Contents:**

**S1. OPGE response of the MLG device**

**S2. Modulated OPGE response for different polarization modulation schemes**

**S3. CPGE extraction for different polarization modulation schemes**

**S4. Response time of the device**

## S1. OPGE response of the MLG device

The derivation of the OPGE response has been fully described in Ref. [1]. In general, the OPGE response arises from the electric quadrupole and magnetic dipole effects induced by the light phase gradient, corresponding to the response term $\boldsymbol{J}_{qp}$. It can be divided into four terms according to its dependence on the SAM and OAM[1]:

$$\boldsymbol{J}_{qp}(\rho,\theta,z) = m\cdot\sigma_i\,\boldsymbol{J}_{(1)}(\rho,\theta,z) + m\,\boldsymbol{J}_{(2)}(\rho,\theta,z) + \sigma_i\,\boldsymbol{J}_{(3)}(\rho,\theta,z) + \boldsymbol{J}_{(4)}(\rho,\theta,z) \quad (S1.1)$$

where the first term $m\cdot\sigma_i\,\boldsymbol{J}_{(1)}$ is proportional to the product of SAM ($\sigma_i$) and OAM (m) and changes its sign when the SAM switches from +1 to -1. Since the SAM is related to the left/right circular polarization and is tunable in circular photogalvanic effect (CPGE) detection, it can be extracted via CPGE measurement, with the extracted component $m\cdot\boldsymbol{J}_{(1)}$ proportional to the OAM order. If we ensure that the total power and ring radius of the OAM beam remain unchanged for different $m$ values, the measured CPGE response from $m\cdot\boldsymbol{J}_{(1)}$ has a quantized magnitude on the OAM order $m$, which enables detection of the OAM order[1-3]. The third term $\sigma_i\,\boldsymbol{J}_{(3)}$ changes its sign when the SAM order $\sigma_i$ switches from +1 to -1, but it shows no dependence on the OAM order $m$ and gives a background signal in the CPGE measurement. The second term $m\,\boldsymbol{J}_{(2)}$ and the fourth term $\boldsymbol{J}_{(4)}$ have no circular polarization dependence and are removed when the circular polarization-dependent component is extracted from CPGE measurements.

Here, we focused on the first term $m\cdot\sigma_i\,\boldsymbol{J}_{(1)}$, which is used for OAM detection. For the multilayer graphene used in this work, symmetry determines that the first term $m\cdot\sigma_i\,\boldsymbol{J}_{(1)}$ has only a radial component with the following expression:

$$m\cdot\sigma_i\,J_{(1)}^{\rho} = m\cdot\frac{4E_0^2|u_{p,m}(\rho,z)|^2}{\rho}\frac{\sigma_i}{1+|\sigma|^2}\left(S_i^{xxyy} - S_i^{xyxy}\right) \quad (S1.2)$$

where $E_0$ is the amplitude of the light field, $u_{p,m}(\rho,z)$ is the normalized LG mode profile, and $\sigma = \sigma_r + i\sigma_i = \tilde{E}_y/\tilde{E}_x$ is the ratio of the complex amplitudes of the light field in two perpendicular polarization directions and describes the arbitrary polarization state of the OAM beam. The specific expression of $u_{p,m}(\rho,z)$ is given in Ref. [1].

For both mechanical and PEM modulation, the polarization undergoes one periodic switch between left circular ($\sigma_i = +1$) and right circular ($\sigma_i = -1$) polarizations in one operation cycle. However, the specific evolution of the polarization sequence (the variation in $\sigma$) is different, which can lead to differences in OPGE signal extraction. Moreover, corresponding to the different polarization modulation schemes, the CPGE extraction approaches are also different for mechanical and PEM modulations, which leads to different constant coefficients for the extracted CPGE response. In Section 2-3, we provide the derivation of the specific analytic expression of the modulated OPGE response and the extraction of the CPGE component to show the differences between these two different polarization modulation schemes.

## S2. Modulated OPGE response for different polarization modulation schemes

To obtain the specific expression of the modulated OPGE response, we derive the expressions of the modulated polarization state, denoted by the complex $\sigma = E_y/E_x$, for mechanical and PEM modulation. For mechanical modulation, a quarter wave plate (QWP) is placed after a linear polarizer with the polarization direction along the x-direction and rotated to modulate the polarization states. When the quarter-wave plate is rotated relative to the polarizer by angle $\theta$, the expression of the light field after the QWP is given by:

$$\tilde{E}_x = E_{p,m}(\vec{r})(\cos^2\theta + i\sin^2\theta), \quad \tilde{E}_y = E_{p,m}(\vec{r})(1-i)\cos\theta\sin\theta \tag{S2.1}$$

and the expression of $\sigma$ is given by:

$$\sigma = E_y/E_x = \frac{(1-i)\cos\theta\sin\theta}{\cos^2\theta + i\sin^2\theta} = \sigma_r + i\sigma_i \tag{S2.2a}$$

$$\sigma_r = \frac{\cos\theta\sin\theta(\cos^2\theta - \sin^2\theta)}{\cos^4\theta + \sin^4\theta}, \sigma_i = -\frac{\cos\theta\sin\theta}{\cos^4\theta + \sin^4\theta}, \frac{1}{1+|\sigma|^2} = \cos^4\theta + \sin^4\theta \tag{S2.2b}$$

If we consider that the QWP is continuously rotated with frequency f and $\theta = 2\pi ft$, the expression of $m \cdot \sigma_i J^\rho_{(1)}$ is given by:

$$m \cdot \sigma_i J^\rho_{(1)} = -m \cdot \frac{2E_0^2 |u_{p,m}(\rho,z)|^2}{\rho}(S_i^{xxyy} - S_i^{xyxy})\sin 4\pi ft = -m \cdot C(\vec{r})\sin 4\pi ft \tag{S2.3}$$

where $C(\vec{r}) = \frac{2E_0^2|u_{p,m}(\rho,z)|^2}{\rho}(S_i^{xxyy} - S_i^{xyxy})$ is the coefficient that is determined by the light field distribution and the rank-4 conductivity tensors of the detection material. Experimentally, if we extract the 180°-periodicity component of the photocurrent response (CPGE component $J_C$), the extracted CPGE response corresponds to $-m \cdot C(\vec{r})$.

In photocurrent measurements, the detected photocurrent can be expressed as the integration of the current density. When the U-shaped electrodes surround a region $S = [R_1, R_2][0, \pi]$ in polar coordinates, the collected CPGE response is given by:

$$I_C^\rho = \int_S -m \cdot C(\vec{r})d\vec{r} = -m \cdot \int_S C(\vec{r})d\vec{r} \tag{S2.4}$$

If we keep the rings of the LG beams embedded inside the region $S$ by adjusting the focal distance, $\int_S C(\vec{r}) d\mathbf{r}$ can be approximated by[1]:

$$\int_S C(\vec{r}) d\mathbf{r} = \int_0^{\pi} d\theta \int_{R_1}^{R_2} \frac{2E_0^2 |u_{p,m}(\rho,z)|^2}{\rho} (S_i^{xxyy} - S_i^{xyxy}) \rho d\rho = 2\pi b_1 (S_i^{xxyy} - S_i^{xyxy})$$
$$\approx \frac{2w_0^2 E_0^2}{(R_1 + R_2)} (S_i^{xxyy} - S_i^{xyxy}) \tag{S2.5}$$

where $b_1 = \int_{R_1}^{R_2} |u_{p,m}(\rho,z)|^2 d\rho \approx \frac{w_0^2}{\pi(R_1+R_2)}$, $w_0^2 = \int_{R_1}^{R_2} |u_{p,m}(\rho,z)|^2 d\mathbf{r}$ remains unchanged for different OAM orders $m$. Therefore, $\int_S C(\vec{r}) d\mathbf{r}$ remains approximately unchanged for different OAM orders, and the collected CPGE response $I_C^\rho$ is proportional to the OAM order m.

For PEM modulation, a polarizer is positioned before the PEM at a 45° angle between the principal axes to produce a 45° linearly polarized beam. A periodic phase retardation $\delta = \delta_0 \sin 2\pi f t$ is created by the PEM in two polarization directions along the principal axes. The expression of the light field after the PEM is given by:

$$E_x = \frac{E_{p,m}(\vec{r})}{\sqrt{2}}, \quad E_y = \frac{E_{p,m}(\vec{r})}{\sqrt{2}} e^{i\delta} \tag{S2.6}$$

Then, $\sigma$ can be written as:

$$\sigma = E_y/E_x = e^{i\delta} = \sigma_r + i\sigma_i \tag{S2.7a}$$

$$\sigma_r = \cos\delta, \sigma_i = \sin\delta, \quad \frac{1}{1+|\sigma|^2} = \frac{1}{2} \tag{S2.7b}$$

The expression of $m \cdot \sigma_i J_{(1)}^\rho$ is as follows:

$$m \cdot \sigma_i J_{(1)}^\rho = m \cdot \frac{2E_0^2 |u_{p,m}(\rho,z)|^2}{\rho} (S_i^{xxyy} - S_i^{xyxy}) \sin[\delta_0 \sin(2\pi f t)]$$
$$= m \cdot C(\vec{r}) \sin[\delta_0 \sin(2\pi f t)] \tag{S2.8}$$

where $\sin\delta = \sin[\delta_0 \sin(2\pi f t)]$ can be expanded by the integral order Bessel function:

$$\sin\delta = \sin[\delta_0 \sin(2\pi f t)] = \sum_{n=0}^{+\infty} 2J_{2n+1}(\delta_0) \sin[2(2n+1)\pi f t] \tag{S2.9}$$

Experimentally, if we extract the component of the photocurrent with frequency f via a lock-in amplifier, the component corresponds to the leading term of $m \cdot \sigma_i J_{(1)}^\rho$ and can be written as:

$$m \cdot \sigma_i J_{(1)}^\rho \approx m \cdot C(\vec{r}) 2J_1(\delta_0) \sin 2\pi f t \tag{S2.10}$$

The extracted CPGE component corresponds to $m \cdot C(\vec{r}) 2J_1(\delta_0)$. Compared with mechanical modulation with a quarter waveplate, CPGE response generation via the PEM modulation approach has an additional $2J_1(\delta_0)$ coefficient, and for $\delta_0 = \pi/2$, $2J_1(\pi/2) = 1.13365$.

### S3. CPGE extraction for different polarization modulation schemes

In the last session, we only consider the difference in the CPGE extraction part because of different polarization modulation schemes; in this session, we calculate the specific expression of the extracted CPGE response under these two measurement schemes because of different CPGE readout approaches.

For PEM modulation, the CPGE response is directly extracted by the lock-in amplifier locked to the 50.14-kHz modulation frequency, and the lock-in amplifier outputs the root mean square of the CPGE response, which can be written as:

$$J_C^{PEM} = m \cdot C(\vec{r}) 2J_1(\delta_0)/\sqrt{2} \tag{S3.1}$$

For mechanical modulation, to exclude background signals and reduce 1/f noise, the OAM beams are also chopped by a mechanical chopper in an on-off manner, and the photocurrent response is extracted by a lock-in amplifier locked to the chopping frequency $f'$. Under this measurement scheme, the photocurrent response to chopped light can be written as:

$$J_{mod} = \begin{cases} m \cdot \sigma_i J_{(1)}^\rho, & \dfrac{k}{f'} \leq t < \dfrac{k+1/2}{f'}, \ k \in Z \\ 0, & \dfrac{k+1/2}{f'} \leq t < \dfrac{k+1}{f'}, \ k \in Z \end{cases} \tag{S3.2}$$

The lock-in amplifier extracts the component with frequency $f'$ and outputs the root mean square of the photocurrent, which is given by:

$$J_{rms} = \frac{f'}{\sqrt{2}} \int_0^{1/f'} J_{mod} \cdot 2\sin(2\pi f' t)\, dt$$
$$= m \cdot C(\vec{r}) \sin 2\theta \cdot \frac{2}{\pi}/\sqrt{2} \tag{S3.3}$$

The extracted CPGE response is given by:

$$J_C^{QWP} = m \cdot C(\vec{r}) \cdot \frac{2}{\pi}/\sqrt{2} \tag{S3.4}$$

For mechanical and PEM modulations, the extracted CPGE response differs with different constant coefficients because of different CPGE measurement schemes. Theoretically, the extracted CPGE response with PEM modulation is $\pi J_1(\pi/2) \approx 1.78$ times that with mechanical modulation.

**S4. Response time of the device**

In this session, we present the characterization of the photocurrent response time of a typical MLG device used in this work. We measure the second-order DC photocurrent under excitation by a basic mode Gaussian beam with no OAM. The beam is electrically modulated at different frequencies (f) (up to 100 kHz) with the controller of the quantum cascade laser, allowing us to evaluate the device's response time, as shown in Fig. S2. The response time ($\tau$) of the device is obtained by fitting with the following function:

$$R(f) = \frac{R_0}{\sqrt{1 + (2\pi f \tau)^2}} \tag{S4.1}$$

The results indicate that the device has a response time of $\tau = 3.42$ μs, which is mainly limited by the RC constant of the device.

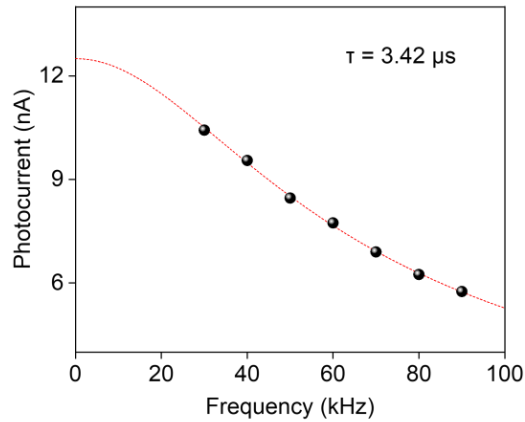

**Fig. S1** Measurements of the response time for modulation by the laser controller